\documentclass{mn2e}
\usepackage{epsf,times}
\newcommand{\etal}{{et al}\/.}

\begin{document}
\title[Jet speeds in quasars]{Testing the beamed inverse-Compton model for
  jet X-ray emission: velocity structure and deceleration?}
\author[M.J.~Hardcastle]{M.J.~Hardcastle\\School of Physics,
  Astronomy and Mathematics, University of Hertfordshire, College
  Lane, Hatfield, Hertfordshire AL10 9AB} \maketitle
\begin{abstract}
By considering a small sample of core-dominated radio-loud quasars
with X-ray jets, I show, as has been argued previously by others, that
the observations require bulk jet deceleration if all of the X-ray
emission is to be explained using the widely adopted beamed
inverse-Compton model, and argue that jets even in these powerful
objects must have velocity structure in order to reconcile their radio
and X-ray properties. I then argue that the deceleration model has several
serious weaknesses, and discuss the viability of alternative models
for the decline in X-ray/radio ratio as a function of position.
Although inverse-Compton scattering from the jets is a required
process and must come to dominate at high redshifts, adopting an
alternative model for the X-ray emission of some nearby, well-studied
objects can greatly alleviate some of the problems posed by these
observations for the beamed inverse-Compton model.
\end{abstract}

\section{Introduction}

\subsection{X-ray jets and the beamed inverse-Compton model}
\label{intro}
X-ray emission from the jets of radio-loud active galaxies is now
known to be common. In the low-power, Fanaroff \& Riley (1974) type I
(FRI) objects, objects that have detectable jets in the radio often
(Worrall, Birkinshaw \& Hardcastle 2001) and perhaps even always
(Hardcastle \etal\ 2002b) have corresponding X-ray jet emission,
although the X-ray/radio ratio varies widely. Because of the
continuity between the radio, optical and X-ray spectra in the
best-studied jets (e.g. Wilson \& Yang 2002; Hardcastle, Birkinshaw \&
Worrall 2001)
the X-ray emission is argued to be synchrotron; the broad-band spectra
of the jets can fitted with one-zone synchrotron models (i.e., models
with a single population of relativistic electrons with $N(E)$
monotonically decreasing as a function of $E$), although neither these
broad-band spectra (e.g. Hardcastle \etal\ 2001) or their detailed
properties (e.g. Hardcastle \etal\ 2003, Perlman \& Wilson 2005) are
consistent with standard continuous-injection models for the particle
acceleration. The X-ray emission must then trace high-energy particle
acceleration (electron Lorentz factors $\gamma \ga 10^7$). It is
coincident, in the best-studied cases, with the region where strong
bulk deceleration is thought to be taking place (Hardcastle \etal\
2002b), and is thus limited to the inner few kpc corresponding to the
one-sided (relativistically beamed) inner part of the jet, although in
at least one case, NGC 6251, X-ray synchrotron emission plausibly
persists out to 100-kpc scales (Evans \etal\ 2005).

It is much less clear that a single mechanism can explain X-ray jet
emission from the more powerful FRII objects. In some low-power FRII
radio galaxies, jet X-rays are plausibly modelled as synchrotron
emission (e.g. 3C\,219, Comastri \etal\ 2003; 3C\,403, Kraft \etal\
2005; Pictor A, Hardcastle \& Croston 2005). But the most widely
studied class of objects to exhibit X-ray jets, following the
discovery of the prototype, PKS 0637$-$752 (Schwartz \etal\ 2000), are
core-dominated quasars (CDQ), and these often have broad-band spectra
in which the optical data points preclude a one-zone synchrotron
model. Nor is it possible to explain the properties of the jets as a
result of inverse-Compton scattering of synchrotron photons
(synchrotron self-Compton, SSC) or cosmic microwave background photons
(CMB/IC) if the jet is non-relativistic and the magnetic field
strength in the jet is close to equipartition, although such models,
with fields close to equipartition, successfully explain the X-ray
emission from other components of powerful radio sources, such as
hotspots and lobes (see Hardcastle \etal\ 2004, Kataoka \& Stawarz
2004, Croston \etal\ 2005, and references therein, for discussions of
these components). Instead, the model proposed independently by
Tavecchio \etal\ (2000) and Celotti \etal\ (2001) is widely adopted.
In this model, the jet is moving relativistically, with a bulk Lorentz
factor $\Gamma$. As seen by the jet, the energy density in the
microwave background increases by a factor of the order $\Gamma^2$,
and this increases the emissivity of the CMB/IC process in the jet
frame. Both because the jet-frame inverse-Compton emissivity is
anisotropic (the CMB is anisotropic in the jet frame) and because of
the strong beaming in the lab frame at high bulk Lorentz factors, this
process is only viable in objects where the jet velocity vector makes
a small angle to the line of sight: however, we know from superluminal
motion observations that this is true of the CDQ population. A key
result of these early analyses of PKS 0637$-$752 was that the angle to
the line of sight, {\it and the bulk Lorentz factor $\Gamma$}, were
consistent with constraints on the pc-scale speed from superluminal
motion observations. Thus the model implied little or no bulk
deceleration between the parsec and 100-kpc scales in these jets, a
point emphasised by Tavecchio \etal\ (2004).

\subsection{Problems with the model}

Since this beamed CMB/IC model (hereafter just referred to as the
CMB/IC model) was first proposed a number of
objections to it have been put forward which to some extent undermine
the original strong arguments for its adoption. These can be
summarized as follows:

\begin{enumerate}
\item Incompatibility with jet speeds from radio observations
\item The inherent unlikelihood of one-zone models
\item Fine-tuning of $\gamma_{\rm min}$
\item Differences between X-ray and radio structures
\item Radio/X-ray ratio changes
\end{enumerate}

I discuss them in detail in the following subsections.

\subsubsection{Jet speeds from radio observations}
\label{intro-jets}
The kpc-scale jet bulk
  speeds required in the CMB/IC model are inconsistent with the best
  constraints available using jet sidednesses and prominences derived
  from radio observations of lobe-dominated radio galaxies and quasars
  (Bridle \etal\ 1994; Wardle \& Aaron 1997; Hardcastle \etal\ 1999;
  Arshakian \& Longair 2004; Mullin \etal , in prep.). The
  best-fitting characteristic speeds derived in these papers (which
  use varying samples and analysis methods) are in the range
  0.5--0.7$c$, and bulk Lorentz factors as high as 10 are definitely
  incompatible with the observations (Mullin \etal , in prep.). As the
  samples used for this sort of work are generally lobe-dominated
  objects (they must be drawn from samples whose orientation to the
  line of sight is unbiased, or at least where the bias is understood,
  in order to allow the statistical inference of the jet speed), one
  could argue that the CDQ are physically different (which would be
  inconsistent with unified models), or that they represent a tail of
  objects with extremely high jet speeds (which would render them
  effectively irrelevant for understanding the physics of radio
  sources in general), but these positions clearly have disadvantages.

  The best way to reconcile the radio observations with the beamed
  CMB/IC model without invoking differences between the population of
  X-ray-jetted CDQ and the population of radio sources in general is
  to suppose that the jet has some velocity structure, in the sense
  that a cross-section of the jet at any given distance from the
  nucleus contains material moving away from the nucleus at different
  speeds. A simple model, for the sake of discussion, would be one in
  which the jet in all powerful sources consisted of a slow outer
  sheath (with $v/c \sim 0.5$) and a fast central spine (with $\Gamma
  \sim 10$). Doppler dimming would then prevent any detection of the
  spine in the lobe-dominated objects used for the radio-based speed
  detections. The spine would dominate at small angles to the line of
  sight, and be responsible for the X-ray jet emission of CDQ via the
  CMB/IC process. Conceivably the sheath could produce X-rays via the
  synchrotron process too, as seen in some lobe-dominated objects (see
  above). The consequences of this spine-sheath model will be
  discussed later in the paper.

\subsubsection{Synchrotron emission and one-zone models}
\label{intro-synch}
The argument for rejecting synchrotron emission as the mechanism for
  the X-rays is based on the assumption that synchrotron emission
  would be described by the kind of one-zone model discussed above (a
  single electron population with monotonically decreasing $N(E)$).
  There are two reasons why arguments based on this assumption should
  be treated with caution. One is the fact (pointed out by Dermer \&
  Atoyan 2002) that $N(E)$ need not monotonically decrease: electron
  losses against the microwave background can produce dips in the
  energy spectrum. The other, simpler point is that we have good
  reasons to suppose that a single electron population {\it cannot}
  describe a kpc-scale region emitting by the X-ray synchrotron
  process. The loss timescale for X-ray-synchrotron-emitting electrons
  in the magnetic field strength appropriate for a jet is of the order
  of tens of years at most, so such electrons can travel at most tens
  of light years from their acceleration site: a spatially uniform
  electron population is therefore not possible unless the
  acceleration mechanism is uniformly distributed. The nearest radio
  galaxy, Cen A, in which the resolution is matched to those spatial
  scales, clearly shows multiple discrete acceleration sites (e.g.
  Hardcastle \etal\ 2003)\footnote{M.\ Lyutikov has pointed out to me
  that this type of fine spatial structure in electron populations is
  also seen in pulsar wind nebulae, such as the Crab (e.g. Bietenholz
  \etal\ 2004).}. If synchrotron emission in more distant objects were
  to follow this pattern, our X-ray flux measurements, with a spatial
  resolution of order $10^3$ times the loss spatial scale, are
  averaging over a number of different electron populations, and the
  resulting effective spectrum need not, and in general will not,
  resemble a one-zone model. As discussed above, the net spectrum of
  the X-ray jet region in FRI radio galaxies {\it is} often reasonably
  well fitted with a one-zone model (one with {\it ad hoc} assumptions
  about the high-energy slope); but we cannot assume that the same
  will be true of more powerful jets, which presumably have very
  different physics and particle acceleration mechanisms.

\subsubsection{Fine-tuning of $\gamma_{\rm min}$}
\label{intro-gamma}

The lowest possible
  Lorentz factor of the electrons plays a crucial role in the CMB/IC model. To
  scatter CMB photons ($T=2.7(1+z)$ K in the frame of the active
  nucleus) into the X-ray (1 keV, lab-frame) requires electrons with
  energies of order $10^3$, since the energy gain in the
  inverse-Compton process goes as $\gamma^2$. But if relativistic
  boosting is important this constraint becomes $\gamma_{\rm min} \sim
  10^3/\Gamma$ (if we make the approximation $\Gamma \sim {\cal D}$,
  where ${\cal D}$ is the Doppler factor). In fact we require
  $\gamma_{\rm min}$ to be significantly less than this, since we
  require the CMB/IC spectrum to peak below 1 keV (otherwise a flat or
  even inverted X-ray spectrum would be observed). At the same time,
  depending on the low-energy electron energy spectrum assumed (and it
  is important to bear in mind that these calculations require an enormous
  extrapolation from the observable electron energies: e.g. Harris
  2004), there may be a {\it lower} limit on $\gamma_{\min}$ from the
  requirement that the CMB/IC prediction should not exceed the
  observed optical flux density of the jet, and there is a definite
  limit on the number of cold ($\gamma \approx 1$) electrons from bulk
  Comptonization of the CMB (Georganopoulos \etal\ 2005) which in the
  case of PKS 0637$-$752 may provide a stringent constraint (Uchiyama
  \etal\ 2005). Depending on the models and assumptions used, these
  different constraints may require very fine tuning of $\gamma_{\rm
  min}$. In addition, the $\gamma_{\rm min}$ values commonly used,
  $\sim 20$, are significantly different from the best estimates of
  $\gamma_{\rm min}$ derived from radio observations of hotspots,
  $\gamma_{\rm min} \sim 500$ (see e.g. Hardcastle 2001).

\subsubsection{X-ray and radio knots}

Tavecchio, Ghisellini \& Celotti
  (2003) pointed out that, since the X-ray emission in the CMB/IC
  model traces very low-energy electrons ($\gamma \sim 10$) we would
  expect it to vary spatially much more smoothly than the radio
  or optical emission, which traces higher-energy electrons with
  shorter radiative lifetimes ($\gamma \sim 10^4$--$10^6$) and which
  also depends on the magnetic field strength. In fact, knots in
  jets such as that in 3C\,273 show similar sizes in the X-ray, radio
  and optical. Tavecchio \etal\ propose to solve this problem by
  suggesting that such knots are composed of many unresolved
  subclumps, but, as pointed out by Stawarz \etal\ (2004), allowing a low
  filling factor makes SSC a viable process again. Similarly, we would
  not expect to see positional offsets between radio and X-ray peaks,
  since a peak in the X-ray should represent an increase in the
  normalization of the electron energy spectrum, and so coincide with
  increased radio brightness, but such offsets are observed in some
  CDQ jets (e.g.
  Siemiginowska \etal\ 2002).

\subsubsection{Radio/X-ray ratio changes}
\label{intro-ratio}
Some quasars show a
  significant change in the radio/X-ray flux ratio as a function of
  distance along the jet, almost always in the sense that there is
  less X-ray emission for a given amount of radio emission further
  from the nucleus. This is seen in samples of objects (e.g. Sambruna
  \etal\ 2004) as well as in individual well-studied objects such as
  3C\,273 (Marshall \etal\ 2001; Sambruna \etal\ 2001). Georganopoulos
  \& Kazanas (2004) have argued, following earlier suggestions, that
  in the CMB/IC model this can be understood as a deceleration of the
  jet. This suggestion will be tested, and its consequences explored,
  in later sections of the present paper.

\subsection{This paper: testing the model}

The list of potential problems presented above makes it clear that a critical
test of the CMB/IC model for quasar jets is urgently needed. Possible
tests fall into two categories: statistical, large-sample tests and
tests that involve the details of particular observations. One
promising approach in the former category is to look for a correlation
between the jet X-ray/radio ratio and the energy density of the CMB,
which increases as $(1+z)^4$. So far no correlation has been found
(Marshall \etal\ 2005) which, in the framework of the CMB/IC model,
implies that there must be a wide range in the jet beaming parameters.
If this is the case, the assembly of a sample large enough to test the
model may take a very long time (and will certainly be dependent on
the award of a large amount of further {\it Chandra} time). Similarly,
it would be desirable to see whether the distribution of angles to the
line of sight required by the CMB/IC model is compatible with the
statistics of the CDQ with respect to their parent population of
unbeamed objects, but at present the CDQ showing X-ray jets are a
heterogeneous population drawn from different parent populations with
different selection criteria, and a test of the model in this way
would probably require new samples to be defined and observed.

As existing observations do not allow a statistical approach, I
concentrate in this paper on the results that can be inferred from
detailed observations of individual objects, and in particular from
the observation that many jets have X-ray/radio flux ratios that are a
strong function of position (Section \ref{intro-ratio}). I construct a
small sample of well-resolved jets and make measurements and infer jet
properties in a systematic way. The results allow some conclusions to
be drawn about the properties of jet speeds if the CMB/IC model is
correct, and motivate further consideration of alternative models.

In what follows I use a concordance cosmology with $H_0 = 70$
km s$^{-1}$ Mpc$^{-1}$, $\Omega_{\rm m} = 0.3$ and $\Omega_\Lambda =
0.7$. Spectral indices $\alpha$ are the energy
indices and are defined in the sense that flux $\propto
\nu^{-\alpha}$. The bulk Lorentz factor of a jet is always denoted
$\Gamma$, while $\gamma$ is used for the random Lorentz factor of electrons.

\section{Sample and analysis}

\begin{table*}
\caption{Properties of the sources in the sample}
\label{props}
\begin{tabular}{llrrrlrr}
\hline
IAU name&Other name&$z$&$S_{1.4}$ (Jy)&$L_{1.4}$ (W Hz$^{-1}$
sr$^{-1}$)&Reference&{\it Chandra} OBSID&Livetime (s)\\
\hline
0518$-$458&Pictor A&0.035&6.3&$1.4 \times 10^{24}$&1, 2&3090&36351\\
0605$-$085&PKS&0.870&1.9&$9.2 \times 10^{26}$&3&2132&8663\\
0827$+$243&B2&0.939&0.9&$5.4 \times 10^{26}$&4&3047&18268\\
1127$-$145&PKS&1.18&5.7&$6.7 \times 10^{27}$&5&866&27358\\
1136$-$135&PKS&0.554&4.3&$6.0 \times 10^{26}$&3, 8&3973&70174\\
1150$+$497&4C\,49.22&0.334&1.6&$5.9 \times 10^{25}$&3, 8&3974&62122\\
1226$+$023&3C\,273&0.158&53.6&$3.2 \times 10^{26}$&6, 7&4876&37457\\
1354$+$195&4C\,19.44&0.720&2.6&$7.4 \times 10^{26}$&3&2140&9056\\
1510$-$089&PKS&0.361&2.7&$1.2 \times 10^{26}$&3&2141&9241\\
\hline
\end{tabular}
\vskip 5pt
\begin{minipage}{14cm}
`IAU name' lists the (B1950-derived) positional name of the source.
  `Other name' lists the name of the source in radio catalogues, if
  one is widely used. `PKS' means that the source is taken from the
  Parkes catalogue and normally referred to with a PKS prefix:
  similarly, `B2' implies a source from the second Bologna catalogue.
  Redshifts are taken from the XJET catalogue, which in turn generally
  takes them from the papers listed in the `Reference' column. 1.4-GHz
  flux densities are taken from FIRST or NVSS maps, except for Pic A, where
  data from Perley \etal\ (1997) are used. Luminosities are calculated
  from the flux densities assuming a spectral index of 0.8. References
  give previously published discussions of the X-ray jet, and
  are as follows: 1, Wilson, Young \& Shopbell (2001); 2, Hardcastle \&
  Croston (2005); 3, Sambruna \etal\ (2004); 4, Jorstad \& Marscher
  (2004); 5, Siemiginowska \etal\ (2002); 6, Marshall \etal\ (2001);
  7, Sambruna \etal\ (2001); 8, Sambruna \etal\ (2005).
\end{minipage}
\end{table*}

The sample of objects analysed here was taken from the XJET catalogue
(http://hea-www.harvard.edu/XJET/), maintained by Harris, Cheung \&
Stohlman, as it stood at June 2005. This catalogue aims to maintain an
up-to-date list of objects for which jet- or hotspot-related X-ray
emission has been claimed in the literature. I selected from this list
FRII objects with X-ray jet detections that are not clearly described
as synchrotron emission, that show extended structure, e.g. an
extended jet or multiple knots, and that have useful radio data
available in the Very Large Array (VLA) archive. The final sample
consisted of 9 objects of which all but one (Pic A) were quasars: Pic
A is of course a broad-line radio galaxy and so probably a low-luminosity
counterpart of a quasar. Although there are good arguments that the
jet in Pic A is synchrotron in orgin (Wilson, Young \& Shopbell 2001;
Hardcastle \& Croston 2005) I retained it in the sample, as it is the
only low-redshift, lobe-dominated source known to show a strong linear
X-ray jet of the type seen in CDQs, and so provides a useful
comparison object. PKS 0637$-$752 itself was excluded because it is not
visible to the VLA. Some basic properties of the sample are given in
Table \ref{props}.

For each object in the sample I retrieved the {\it Chandra} X-ray data
for from the public archive. In some cases there was more than one
observation, and here generally I simply used the longest: the aim
here is to have sufficient statistics to make a flux measurement,
rather than to make the best possible X-ray maps. The {\it Chandra}
observation IDs and corresponding livetimes used are tabulated in
Table \ref{props}. X-ray analysis was carried out using {\sc ciao}
3.2.1 and CALDB 3.0.0.

VLA data were used to obtain single-frequency radio maps with a
resolution as closely matched to that of {\it Chandra} as possible.
The images for 0605$-$085, 1136$-$135, 1150$+$497, 1354$+$195 and
1510$-$089, which were presented in Sambruna \etal\ (2004), were
kindly made available to me by C.\ C.\ Cheung and are now available
from the XJET web page. R.\ A.\ Perley provided me with images of
Pic~A from Perley, R\"oser \& Meisenheimer (1997). I retrieved data
for 0827$+$243, 1127$-$145 and 3C\,273 from the public VLA archive
[the first two datasets being the basis of the images presented in
Jorstad \& Marscher (2004) and Siemiginowska \etal\ (2002)
respectively]. Where I retrieved VLA data from the archive myself,
they were reduced in the standard manner using AIPS, with several
iterations of self-calibration typically being used to improve the
dynamic range.

\begin{figure*}
\epsfxsize 16.5cm
\epsfbox{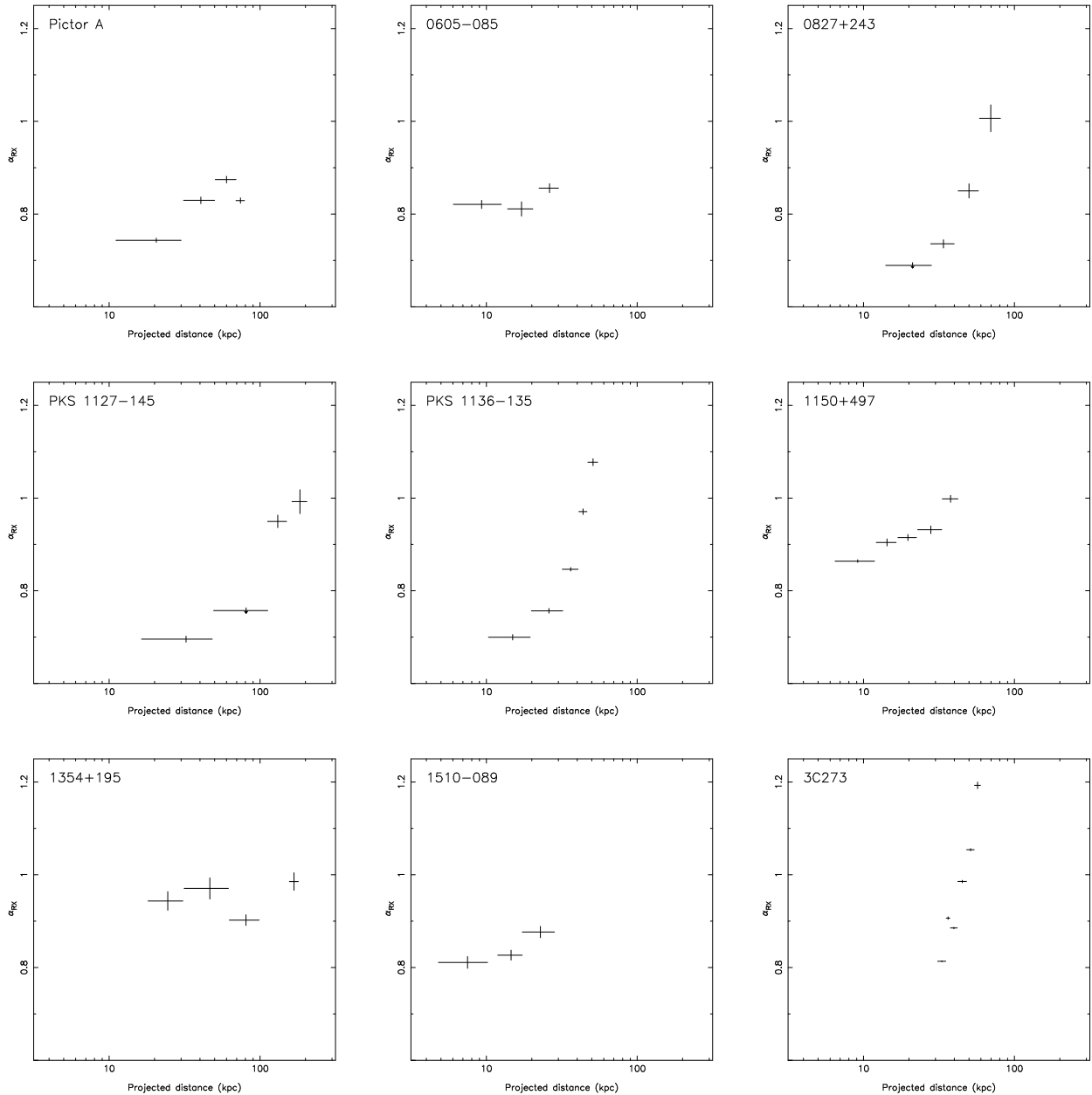}
\caption{The two-point radio/X-ray spectral index of the jet
  components as a function of projected distance. Arrows indicate
  upper limits on $\alpha_{\rm RX}$, resulting from upper limits on the
  radio emission.}
\label{alpha}
\end{figure*}

Once X-ray data and radio maps were available, I divided each jet into
a number of sub-regions, taking radio and X-ray measurements for each.
The regions were defined in general on the basis of the X-ray images,
since the separation into sections was limited by the X-ray signal-to-noise,
but their dimensions were measured from the radio data. Because the
number of counts from any given jet was small, I fitted a power-law
spectrum with Galactic absorption to the whole jet for each source and
used this spectrum to determine the conversion between counts in the
0.5--5.0 keV energy range and flux density. As this conversion factor
is only a weak function of power-law index, this procedure should not
introduce much uncertainty into the measurements. X-ray counts and radio flux
densities were then measured from identical regions, with background
subtraction where necessary (always for X-ray but only occasionally
for radio). In a couple of cases there was no significantly detected
radio emission from an X-ray region, and in this case an upper limit
was determined from the off-source noise on the radio map. The X-ray
counts were converted to a 1-keV flux density using the conversion
factors determined from the spectral fitting. The length
of the region, the typical jet radius within the region, and the
distance of the region from the nucleus (measuring along the jet to
the centre of the region) were also recorded. All these data for each
region are tabulated in Table \ref{regions}. A plot of the radio/X-ray
spectral index, $\alpha_{\rm RX}$, derived from these measurements is
shown in Fig.\ \ref{alpha}; this illustrates the point that
$\alpha_{\rm RX}$ generally increases with distance along the jet.

The radio and X-ray data were then modelled using a version of the
code described by Hardcastle \etal\ (2002a) which numerically
integrates the inverse-Compton equations set out by Brunetti (2000) in
the rest frame of the jet, and then transforms back to the lab frame.
This avoids the use of some of the analytical approximations adopted
in the literature, which are strictly correct only in the limit
$\Gamma \gg 1$. Each jet segment is taken to be a homogenous cylinder
with length and width determined by the measurements tabulated in
Table \ref{regions}. The modelling takes account of the fact that the
inferred volume of the jet, which has a significant effect on the
inverse-Compton emission, is dependent on the angle to the line of
sight $\theta$. I follow Scheuer \& Readhead (1979) in adopting $S =
S'{\cal D}^{(2+\alpha)}$ as the appropriate Doppler boosting formula
for continuous jets emitting isotropically in their rest frames, and I
assume that the rest-frame synchrotron emission is isotropic. I
further assume, as is conventional, equipartition between electron
energy and magnetic field, so that the radio data point provides the
normalization for the electron energy spectrum, and that there are no
energetically significant protons\footnote{The reason for the
assumption of no energetically significant protons is that, as argued
by Hardcastle \etal\ (2004) and Croston \etal\ (2005) for hotspots and
lobes respectively,
the similarity between the energy densities in the magnetic fields
measured in the large-scale components of powerful radio sources and
the energy densities in the electron population has to be a
coincidence if there is an energetically dominant proton population.
Since the kpc-scale jets directly feed the lobes and hotspots, this
seems a reasonable inference.
If there is a population of protons or other non-radiating particles
with an energy $\kappa$ times that of the electrons, and the magnetic
field is in equipartition with the total energy density of the
particles, then the inverse-Compton emissivity goes as
$(1+\kappa)^{-{{p+1}\over{p+5}}}$, where $p$ is the electron energy
index discussed in the text: thus moderate contributions from
protons, with $\kappa \sim 1$, make little difference to the
results.}. The electron energy spectrum is assumed to be a power law
with index $p$ between electron Lorentz factors $\gamma_{\rm min}$ and
$\gamma_{\rm max}$. As discussed above (Section \ref{intro-gamma}),
$\gamma_{\rm min}$ has to be $\sim 10$ or lower to allow the CMB to be
scattered by the jet and to avoid a flat or inverted spectrum in the
X-ray. $\gamma_{\rm max}$ is less important to the result so long as
it corresponds to frequencies well above the lab-frame radio
constraints. $p$ would in principle be constrained by multi-frequency
radio observations, but few of these are available. In what follows I
assume $p=2.5$ (corresponding to a radio spectral index $\alpha_{\rm
R} = 0.75$), $\gamma_{\rm min} = 10$, and $\gamma_{\rm max} = 4 \times
10^5$. I have verified that the choices of parameters do not have a
strong effect on the qualitative results presented here, though
clearly some changes can have a significant quantitative effect.

\begin{table}
\caption{Measurements from the regions used in the paper}
\label{regions}
\begin{tabular}{lrrrrrr}
\hline
Source&Freq.&Distance&Length&Width&Radio&X-ray\\
&(GHz)&(arcsec)&(arcsec)&(arcsec)&flux&flux\\
&&&&&(mJy)&(nJy)\\
\hline
Pictor A&1.4&29.8&27.4&2.0&6.5&5.0\\
&&58.9&27.4&2.0&16.8&2.5\\
&&86.9&27.4&2.0&38.7&2.5\\
&&107.6&14.0&2.0&18.4&2.8\\[2pt]
0605$-$085&4.9&1.4&1.0&0.3&11.1&5.3\\
&&2.6&1.0&0.3&4.8&2.7\\
&&4.0&1.2&0.6&19.1&4.9\\[2pt]
0827$+$243&4.9&3.2&2.1&0.5&$<$0.5&2.7\\
&&5.1&1.8&0.5&1.0&2.1\\
&&7.6&2.3&0.5&3.7&1.1\\
&&10.5&3.3&0.5&14.1&0.3\\[2pt]
1127$-$145&8.5&4.8&4.7&0.5&0.4&2.7\\
&&11.9&9.3&0.6&$<$0.7&1.6\\
&&19.3&5.5&0.7&8.5&0.7\\
&&27.0&6.2&0.8&10.4&0.4\\[2pt]
1136$-$135&4.9&2.6&1.6&0.5&0.3&1.3\\
&&4.6&2.2&0.5&1.2&1.8\\
&&6.4&1.5&0.5&10.3&3.1\\
&&7.7&0.9&0.5&38.3&1.3\\
&&9.0&1.4&0.6&165.7&0.8\\[2pt]
1150$+$497&4.9&2.1&1.2&0.2&28.4&6.3\\
&&3.2&1.0&0.2&12.5&1.4\\
&&4.5&1.3&0.2&16.0&1.5\\
&&6.3&2.3&0.2&15.9&1.1\\
&&8.6&2.0&0.9&50.5&1.0\\[2pt]
3C\,273&4.8&12.6&1.6&0.2&95.0&51.2\\
&&13.9&0.8&0.7&77.0&7.9\\
&&15.2&1.6&0.8&207.3&31.2\\
&&17.3&2.3&1.2&549.4&14.0\\
&&19.6&2.2&1.2&1979.6&14.9\\
&&21.8&2.0&1.2&3138.3&2.0\\[2pt]
1354$+$195&4.9&3.9&2.0&0.2&17.3&0.9\\
&&7.5&4.9&0.2&33.4&1.1\\
&&13.0&5.8&0.2&22.3&2.5\\
&&27.0&3.7&1.5&50.8&1.3\\[2pt]
1510$-$089&4.9&1.6&1.2&0.1&6.9&3.9\\
&&3.2&1.2&0.2&8.3&3.6\\
&&4.9&2.4&0.3&18.0&3.2\\[2pt]

\hline
\end{tabular}
\end{table}

\section{Sample results}

\subsection{Angles}

There is a degeneracy between the angle to the line of sight $\theta$
and the bulk Lorentz factor $\Gamma$ required to produce a given
amount of X-ray emission from the observed radio. The approach I take
here is therefore to keep one fixed while varying the other (cf.
Marshall \etal\ 2005). Ideally, we would know $\theta$ for a given
source independently, but for most of these sources there are no
useful constraints. I began by keeping $\Gamma$ fixed ($\Gamma = 10$,
as used by Marshall \etal) and solving for the required $\theta$ for
each source and jet region. This showed that small angles to the line
of sight ($<10^\circ$) are generally required, particularly at the
bases of the jets, if the CMB/IC process is to account for the
observed X-rays with plausible bulk Lorentz factors. The angle to the
line of sight required for a fixed $\Gamma$ increases systematically
along the jets, as we would expect from Fig.\ \ref{alpha}: if this
were true, it would imply that the jets were bending away from the
line of sight. While this is not impossible -- selection effects
require that the jet be aligned close to the nucleus, but say nothing
about what it should do further out -- there is in general no evidence
in the maps for jet bending of similar magnitude (tens of degrees) in
the plane of the sky (though one source, 0827$+$243, does have a sharp
bend of nearly 90$^\circ$ in projection). So a model in which angle
changes are responsible for the change in radio/X-ray ratio seems
implausible.

\subsection{Lorentz factors}

\begin{figure*}
\epsfxsize 16.5cm
\epsfbox{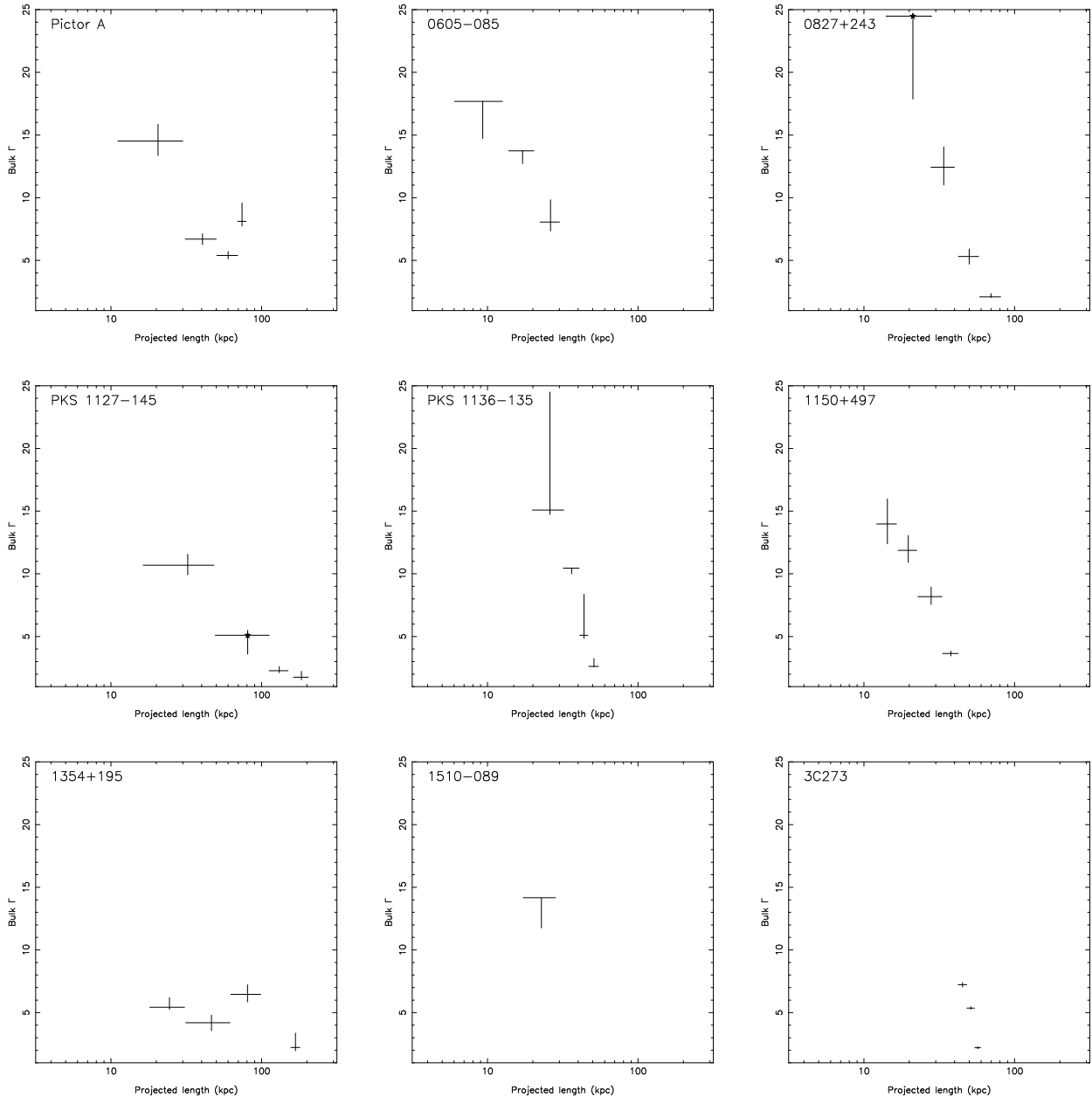}
\caption{Inferred bulk Lorentz factor $\Gamma$ as a function of
  projected length along the jet for the 9 sample sources, assuming
  $\theta = 4^\circ$.}
\label{gamma-r}
\end{figure*}

The more intrinsically probable model to test is the one in which the
speed of the jet changes as a function of distance. I used the results
of the previous subsection to choose a fixed angle, $\theta =
4^\circ$, that represents a compromise between extreme projection and
the requirement that the CMB/IC model be able to produce the observed
X-rays, and solved for $\Gamma$ for each source component. The results
are plotted for each source in Fig.\ \ref{gamma-r}. Errors are the
statistical errors only, derived from the $1\sigma$ (Poisson) errors
on the measured jet X-ray flux densities, and no attempt is made to
represent the (potentially large) systematic uncertainties due to
inadequacies of the model. Some components are not plotted because a
$4^\circ$ angle to the line of sight does not allow the model to
produce the observed X-rays from the measured radio emission: thus,
for example, PKS 1510$-$089 only has one data point on the plot,
corresponding to the outermost region of the jet. Similarly, some
upper error bars are not plotted because the model cannot reproduce
the whole $1\sigma$ range possible from the X-ray data. In some cases
the values I plot in Fig.\ \ref{gamma-r} are significantly different
from published values for the sources in question: I attribute this to
a combination of a different approach (the use of a fixed $\theta$, as
opposed to trying to determine it for each source), a different method
(as discussed above) and, in a number of cases, different radio and/or
X-ray data.

The trend seen in the plot for almost every individual source is
clearly in the sense that lower bulk Lorentz factors are required at
larger distances from the nucleus. The inner parts of the jet often
require $\Gamma \ga 15$ while the outer parts typically have $\Gamma
\la 5$. Although the detailed numbers will depend on the actual angle
to the line of sight, I verified using other values that the trend
with distance does not.

\subsection{Energy transport}
\label{transport}

\begin{figure*}
\epsfxsize 16.5cm
\epsfbox{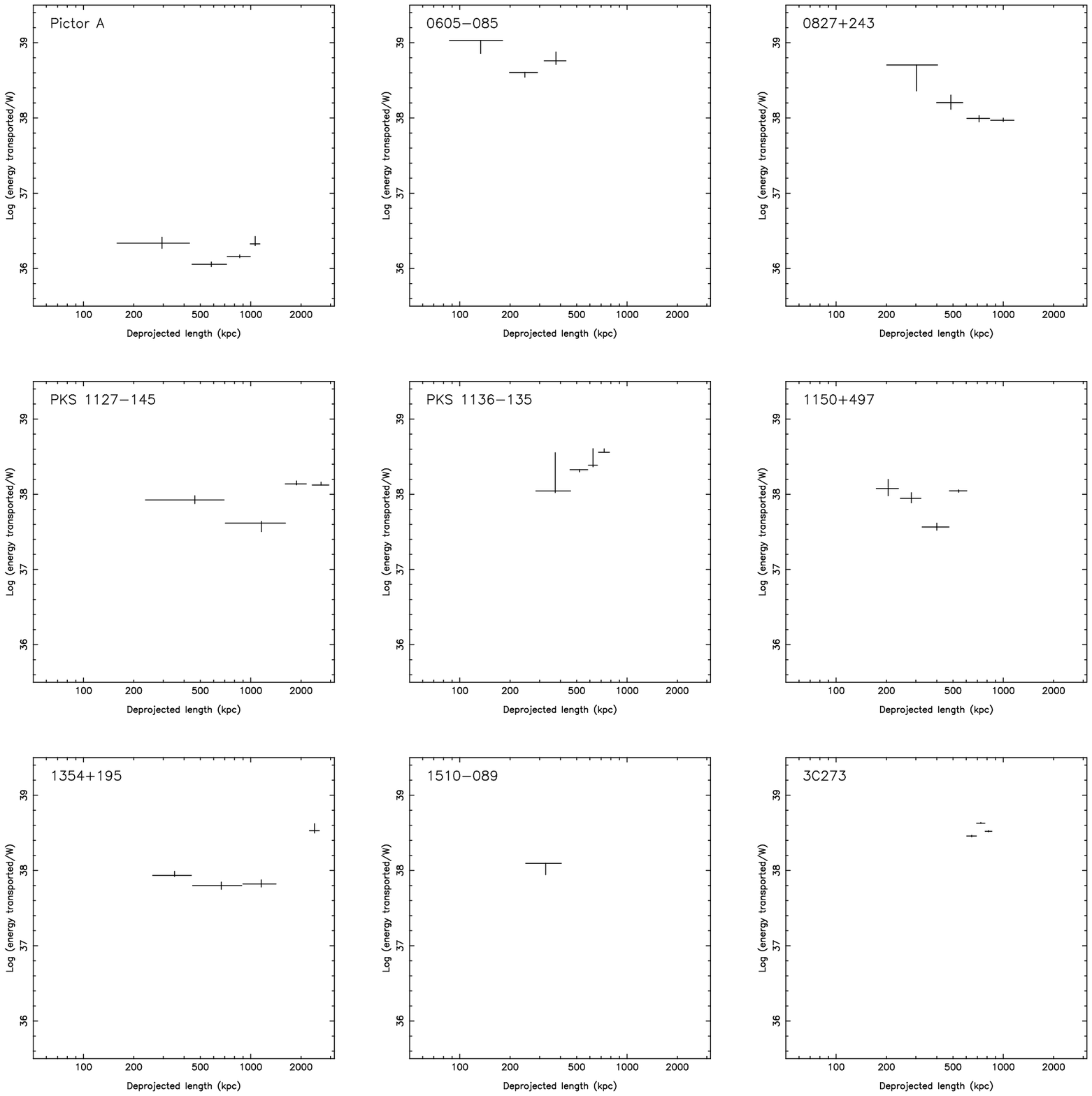}
\caption{Inferred jet power (energy transport along the jet) as a
  function of deprojected distance along the jet assuming
  $\theta = 4^\circ$.}
\label{power}
\end{figure*}

Since the rest-frame energy density $U$ and speed of the jet are
determined by the fits, the lab-frame energy carried by the jet in the
form of magnetic field and electrons, $W = Ac\Gamma^2U$, can be
calculated, where $A$ is the cross-sectional area and $c$ the speed of
light. The results of this calculation are plotted in Fig.\
\ref{power}: as in the previous section, $\theta = 4^\circ$ is
assumed. The value of $W$ for a given jet component (unlike the
Lorentz factor or angle) have a strong dependence on the assumed jet
radius, which is hard to measure accurately. However, it is
interesting that in many cases the power is approximately constant,
within a factor of a few, which is certainly not the case for the
quantity $\Gamma^2$; this is consistent with what we might expect if
the CMB/IC+deceleration model were correct. The powers implied by this
calculation, which of course assumes no protons are present in the
jet, are of the order of $10^{38}$ -- $10^{39}$ W (except in the case
of the nearby Pic A), which are reasonable (cf.\ the jet powers
derived on similar minimum-energy assumptions by Rawlings \& Saunders
1991). They are comparable to the `luminosities' estimated for the
X-ray jets, but of course the emission from these is not isotropic in
the CMB/IC model and so the true lab-frame luminosity is much lower.
Except for the sources where there is a big change in the energy
requirements along the jet, e.g. at the end of 1354$+$195, there is
little here to provide evidence against the CMB/IC model. Note that
the angle to the line of sight adopted means that some of the sources
would be required to be very large. These sources (1127$-$145 and
1354$+$195 in particular) probably in reality have larger angles to
the line of sight and would require correspondingly larger $\Gamma$.

Protons in the jets could well dominate the jet energetics while still
being few enough, or cold enough, not to violate the assumptions set
out in Section 2 about the nature of the internal energy density of
the jet. There is no way of estimating the number of cold protons
present in the jet but, to take one possible scenario, if there were
the same number density of cold protons and energetic electrons, the
energy transported would increase by two orders of magnitude. One
possibly interesting point to make is that if protons dominate the
energy transport, then we might expect the quantity plotted in Fig.\
\ref{power} to increase with distance along the jet, as the jet
decelerates and the bulk kinetic energy of the protons is translated
into internal energy. If the deceleration model is correct, then the
fact that this is not observed suggests that protons do not greatly
dominate the energy transport on these scales.

\section{Discussion}

\subsection{Deceleration...}

As the previous sections have shown, deceleration of jets, inferred by
e.g. Georganopoulos \& Kazanas (2004), is required by the data in the
context of the simplest CMB/IC model, in which jets are homogeneous
cylinders with uniform velocities. The deceleration takes place on
scales of hundreds of kpc, comparable to the size of the radio lobes
if we assume that these sources are normal classical double radio
galaxies seen in extreme projection.

However, we already know from the speeds inferred from properties of
kpc-scale jets (Section \ref{intro-jets}) that the most likely
situation is that jets do not have uniform velocities. One of the most
surprising implications of the deceleration model is that the fast
spine of the jet (producing the X-rays in the CMB/IC model) must
decelerate from $\Gamma > 15$ to $\Gamma < 5$ on 100-kpc scales
without there being any evidence for such deceleration in the slower
outer regions. No tendency is observed in studies of FRII sources
(e.g. Hardcastle \etal\ 1997, Gilbert \etal\ 2004, Mullin \etal\ in
preparation) for the jets to become more two-sided, or more prominent,
at large distances (at least until the jet enters the area of the
hotspots). It is hard to see how, in this two-speed model of the jets,
the slower material can retain its speed while the faster material
decelerates, at least in any picture in which the deceleration is the
result of some influence from outside the jet: the same is true of any
more sophisticated model in which there is a continuous range of
speeds in the jets. A detailed model of the deceleration process must
explain this observation.

Another important question is the mechanism for deceleration. In FRI
jets, the evidence for jet deceleration is overwhelming, in view of
the transition between one-sided and two-sided jets on scales of a few
kpc (see e.g. Laing \& Bridle 2002). The physical mechanism invoked
for FRIs, in order to decelerate the jets while conserving their
momentum, is entrainment of external material. It has been argued
(e.g. Bowman, Leahy \& Komissarov 1996) that a relativistic jet can
entrain matter and undergo substantial deceleration without
catastrophic dissipation of the energy flux of the jet (which we know
cannot be the case in the quasars both from the results of Section
\ref{transport} and from the fact that the jets do not generally
become much more luminous at large distances). Bowman \etal\ argue
that relativistic jets will not entrain efficiently via turbulence at
a boundary layer, as they do in Bicknell's (1984) model of trans-sonic
FRI jets: instead, they propose that mass loss from stellar winds into
the jet is sufficient. This cannot be the case on the scales we are
concerned with, however, for the deceleration is important on scales of
hundreds of kpc, well beyond the scale at which there is a significant
density of stars from the host galaxy. Setting this aside, let us
suppose that entrainment injects cold, dense material into the jet.
Momentum flux balance then would imply that the final bulk Lorentz
factor would be controlled by the density of entrained material: if
the jet is initially entirely composed of a relativistic fluid then we
have, requiring momentum flux to be constant\footnote{Here we are
  assuming that the jet propagates through a constant-pressure cocoon,
  so that the term that arises due to the external pressure gradient
  (Bicknell 1994) may be neglected.},
\begin{equation}
{\Gamma_i\over\Gamma_f} = \sqrt{{A_f\over A_i}\left({U_f + \rho
    c^2}\over{U_i}\right)}
\end{equation}
where subscripts $f$ and $i$ refer to the final and initial states
respectively, and $\rho$ is the density of entrained material in the
final jet. For significant deceleration the rest-frame energy density
in the entrained material must be comparable to, or exceed, the energy
density in relativistic particles, and this would require entrained
masses of $10^3$--$10^4M_\odot$ in regions at the end of the jet in
the most extreme cases, which is certainly not possible by stellar
wind entrainment, where only around $1M_\odot$ is plausibly available
(using mass loss rates from Bowman \etal\ 1996). If protons dominate
the jet energy/momentum flux, the masses required are clearly higher.
Some sort of boundary-layer entrainment might be possible, but runs
into two further problems. Firstly, the energy densities of the
relativistic material do not appear to decrease significantly, as
implied by the results of Fig.\ \ref{power}: if a non-relativistic
component were coming to dominate the momentum and so kinetic energy
density of the source, the energy transported by relativistic
particles should decrease markedly, but it does not. Secondly, any
boundary-layer process should presumably be particularly efficient in
coupling the slow outer layers of the jet with the fast inner spine,
and produce increasing two-sidedness of the kpc-scale jet in the more
weakly beamed FRII radio galaxies and lobe-dominated quasars, which,
as discussed above, is not observed.

The relative constancy of the energy transported by relativistic
particles and field (Fig.\ \ref{power}) in fact suggests that the
deceleration, if real, takes place by translating bulk kinetic energy
into internal particle energy. One process that is worth considering,
because it does not require any particular coupling between the
slow and fast regions of the jet,
is internal shocks in a jet composed of multiple sub-regions of
different speeds (e.g.\ Spada \etal\ 2001). These shocks
cause the Lorentz factor of the jet to tend to a constant
value, rather than decreasing monotonically along the jet, but beaming
effects mean that we would tend to see the inner parts of the jet
as dominated by faster-moving material. However, in order to produce
Lorentz factors as low as those calculated for the outer parts of the
jet, the process generating the different regions (`shells', in the
terminology of Spada \etal) would have to produce some with very low
$\Gamma$, and considerable fine tuning would then seem to be required
to make sure that the interactions happened on 100-kpc scales rather
than the sub-pc scales discussed by Spada \etal.

One positive point in favour of a deceleration model in the CMB/IC
framework is that it helps to explain why Marshall \etal\ (2005) see
little correlation between CMB/IC emission and $(1+z)$. If jets
decelerate, then the variation in $\Gamma^2$ alone is enough to wash
out much of the dependence on rest-frame CMB energy density, setting
aside other important sources of scatter such as source geometry and
angle to the line of sight. On the other hand, permitting deceleration
removes what was initially an attractive feature of the CMB/IC model, namely
the apparent similarity between the jet speeds on pc and kpc scales
(cf.\ Tavecchio \etal\ 2004).

\subsection{... or something else?}

As the discussion above shows, there are some potentially serious
problems for a deceleration model in the CMB/IC framework. Can any
other model explain the observations?

If we wish to retain an inverse-Compton origin for the X-ray emission,
then one obvious possibility is to relax the assumption of
equipartition of energy between the electrons and magnetic field.
Although there is a good deal of evidence that the magnetic field
strengths in the hotspots and lobes of FRII sources are close to the
equipartition values [see, respectively, Hardcastle \etal\ (2004) and
Croston \etal\ (2005) for discussions of the X-ray emission from these
components] the inferred field strengths are generally slightly below
the equipartition value. More importantly, as the mechanism by which
the electrons and magnetic fields come into equipartition is not
clear, there is no very strong reason to suppose that jets follow the
pattern of lobes and hotspots. For the assumed model, a magnetic field
that is lower than the equipartition value by a factor 2 (in terms of
magnetic field strength) corresponds to an increase in the X-ray flux
density of a factor $\sim 3.5$, or a change of $0.07$ in radio/X-ray
spectral index (Fig.\ \ref{alpha}): comparatively small departures
from equipartition can have large effects on the observed X-rays. The
obvious difficulty with this model is that there is no {\it a priori}
reason for supposing that the magnetic field strength will vary in the
required way. Nor is it obvious how this model could be tested. If we
relax the assumption of equipartition, then we can produce the X-ray
emission with bulk speeds much closer to the speeds estimated from the
large-scale radio jet properties. Of course, with large enough
departures from equipartition, boosting of the CMB using large bulk
Lorentz factors is not necessary at all.

The most obvious alternative model is that the X-ray emission is, at
least partially, not inverse-Compton emission at all, but synchrotron
radiation. Although CMB/IC is a required process, and must eventually
come to dominate with increasing redshift, whether or not the bulk
Lorentz factor is high, that does not necessarily imply that it is the
process we see in the (mostly relatively nearby) objects discussed
here. As I discussed in Section \ref{intro-synch}, the arguments
against synchrotron models are not yet strong, and indeed synchrotron
models have already been adopted for some of the bright inner knots of
objects in the current sample (Sambruna \etal\ 2004), while there is
an active controversy over the origin of the X-rays from 3C\,273
(Marshall \etal\ 2001, Sambruna \etal\ 2001). Jester \etal\ (2002)
have shown that the spectrum of components of the 3C\,273 jet turns up
in the infrared-ultraviolet wavelength range, so that a second
spectral component is unambigously required to be present:
qualitatively, this is consistent either with a synchrotron model or
with an inverse-Compton model with a suitably low $\gamma_{\rm min}$.
We know (Section \ref{intro}) that synchrotron emission is possible
from the jets of FRII sources, though it should be noted that at
present there is no unambiguous detection of synchrotron emission from
the jet of a source comparable in luminosity to the core-dominated
quasars in the present sample. Again, we have no detailed physical
understanding of what might cause a change in the amount of X-ray
synchrotron emission produced as a function of distance along the jet.
However, it is worth noting that a very similar trend is seen in many
FRI radio sources, and in such sources the jet emission is certainly
synchrotron in origin. Occam's razor suggests that we should think
very carefully about invoking two entirely different processes to
describe such similar phenomena.

\section{Summary and conclusions}

I have analysed a small heterogeneous sample of objects with known
extended X-ray jets in the framework of the boosted CMB
inverse-Compton (`CMB/IC') model originally set out by Tavecchio
\etal\ (2000) and Celotti \etal\ (2001). Assuming that the CMB/IC
process accounts for all the X-rays, that the jets can be modelled as
homogenous cylinders, that the radio emission determines the geometry,
and that equipartition between electrons and magnetic fields holds, I
find that significantly lower speeds are required at larger distances
from the nucleus in almost all of the sources studied, consistent with
earlier suggestions (e.g.\ Georganopoulos \& Kazanas 2004) but
distinctly inconsistent with the idea that the parsec-scale speed is
reproduced on the kpc scale (e.g.\ Tavecchio \etal\ 2004). Although the
quantitative estimates of $\Gamma$ presented are based on the
assumption of a single angle to the line of sight, the qualitative
picture is independent of that assumption.

From these observations the following conclusions can be drawn:

\begin{itemize}
\item The speeds required (in the objects studied here, and in
  general) are much larger than the speeds inferred from studies of
  the kpc-scale jets in lobe-dominated objects: thus, if the CMB/IC
  model is correct, jet velocity structure is almost certainly required.

\item If the underlying assumptions are true, then deceleration is
  required, at least in some sources: but it is far from clear how
  the deceleration can take place, and there is no evidence for
  deceleration in the properties of lobe-dominated objects, which
  might be expected in a model where jet velocity structure
  was present.

\item If deceleration is to be avoided, some other process must
  account for the decreasing X-ray/radio ratio along the jets. If the
  equipartition assumption is relaxed, then the observations can be
  modelled (admittedly in an entirely {\it ad hoc} way) in terms of a
  varying magnetic field/electron energy density ratio. If the
  assumption that CMB/IC is the only emission process is relaxed, and
  synchrotron radiation provides some or all of the X-rays, then the
  jets of the core-dominated quasars studied here have properties
  strikingly similar to those of the much lower-power FRI radio
  galaxies.
\end{itemize}

\section*{Acknowledgements}

An early version of this work was presented at the Banff meeting on
Ultra-Relativistic Jets in Astrophysics (2005 July 11--15). I am
grateful to the participants there, particularly Dan Harris, Robert
Laing, Maxim Lyutikov, Hermann Marshall, Eric Perlman and Dan
Schwartz, for constructive discussions of the physics of quasar jets,
and to Judith Croston for helpful comments on the first draft of the paper. Any
controversial or erroneous statements remaining in the paper are my
responsibility, not theirs! Teddy Cheung kindly provided me with radio
maps for some of the quasars in the sample and Rick Perley provided
the maps of Pictor A. I am grateful to an anonymous referee for
constructive comments that helped me to improve the paper. The
National Radio Astronomy Observatory is a facility of the National
Science Foundation operated under cooperative agreement by Associated
Universities, Inc.

\clearpage
\end{document}